\documentclass[aps, prb, preprint, showpacs, amssymb, amsmath, superscriptaddress]{revtex4-1}
\usepackage{mathrsfs}
\usepackage{bm}
\usepackage{times}
\usepackage{graphicx}
\usepackage[colorlinks,citecolor=blue,linkcolor=blue]{hyperref}
\usepackage{array}
\usepackage{float}
\usepackage{colortbl}
\usepackage{multirow}
\usepackage{tabularx}
\begin{document}
\title{Robust large-gap topological insulator phase in transition-metal chalcogenide ZrTe$_4$Se}

\makeatletter
\renewcommand*{\@fnsymbol}[1]{\ensuremath{\ifcase#1\or *\or \dag\or \dag \or
    \mathsection\or \mathparagraph\or \|\or **\or \dag\dag
    \or \dag\dag \else\@ctrerr\fi}}
\makeatother
\author{Xing Wang}
\affiliation{State Key Laboratory of Metastable Materials Science and Technology $\&$ Key Laboratory for Microstructural Material Physics of Hebei Province, School of Science, Yanshan University, Qinhuangdao 066004, China}
\affiliation{College of Science, Hebei North University, Zhangjiakou 07500, China}
\author{Wenhui Wan}
\affiliation{State Key Laboratory of Metastable Materials Science and Technology $\&$ Key Laboratory for Microstructural Material Physics of Hebei Province, School of Science, Yanshan University, Qinhuangdao 066004, China}
\author{Yanfeng Ge}
\affiliation{State Key Laboratory of Metastable Materials Science and Technology $\&$ Key Laboratory for Microstructural Material Physics of Hebei Province, School of Science, Yanshan University, Qinhuangdao 066004, China}
\author{Yong Liu}
\email{yongliu@ysu.edu.cn,or ycliu@ysu.edu.cn}
\affiliation{State Key Laboratory of Metastable Materials Science and Technology $\&$ Key Laboratory for Microstructural Material Physics of Hebei Province, School of Science, Yanshan University, Qinhuangdao 066004, China}
\date{\today}


\begin{abstract}
  Based on density functional theory (DFT), we investigate the electronic properties of bulk and single-layer ZrTe$_4$Se. The band structure of bulk ZrTe$_4$Se can produce a semimetal-to-topological insulator (TI) phase transition under uniaxial strain. The maximum global band gap is 0.189 eV at the 7\% tensile strain. Meanwhile, the Z$_2$ invariants (0; 110) demonstrate conclusively it is a weak topological insulator (WTI). The two Dirac cones for the (001) surface further confirm the nontrivial topological nature. The single-layer ZrTe$_4$Se is a quantum spin Hall (QSH) insulator  with a band gap 86.4 meV and Z$_2$=1, the nontrivial metallic edge states further confirm the nontrivial topological nature.  The maximum global band gap is 0.211 eV at the tensile strain 8\%. When the compressive strain is more than 1\%, the band structure of single-layer ZrTe$_4$Se produces a TI-to-semimetal transition.  These theoretical analysis may provide a method for searching large band gap TIs and platform for topological nanoelectronic device applications.
  \end{abstract}
\pacs{61.82.Ms, 73.20.At, 71.20.-b, 73.43.-f}

\maketitle
\section*{I. Introduction}
The TIs are a new quantum state matter with gapped bulk band and gapless edge state, and the low-energy scattering of the edge states leads to dissipationless transport edge channels. The two dimension (2D) TI also called the QSH insulator was first theoretically predicted in 2006 and experimentally observed in HgTe/CdTe quantum wells\cite{HgTe1,HgTe2}. The three dimension (3D) TI was first predicted and observed in the Bi$_{1-x}$Sb$_x$ alloy\cite{Fu2006,bisb}. These pioneering works opened up the exciting field of TIs, expanding at a rapid pace. In the past decade, the more and more compounds have been predicted to be TIs \cite{TI1,TI2,TI3,TI4,TI5,TI6,TI7,TI8,TI9,TI10,TI11}, which has undoubtedly a dramatic impact on the condensed matter physics. However, the extremely small bulk band gaps hinder their applications due to weak spin-orbit coupling (SOC). Therefore, the researchers have strong motivation for exploring new TIs or transforming materials into TIs with large band gaps. \\

 \indent Nowadays, the ZrTe$_5$ has attracted broad attention because of their topological properties\cite{zrte1,zrte2,zrte3,zrte4,zrte5,zrte6,zrte7}. The 3D crystal is located near the phase boundary between strong and weak TIs, the 2D is predicted to be a QSH insulator\cite{zrte1}. Later studies indicate that the topological nature in this bulk material is very sensitive to the crystal lattice constants and detailed composition\cite{zrte8,zrte9}, so it is ideal platforms to investigate different intriguing physical properties. In addition, homologue substitution provides a useful guess for a novel material\cite{2014Computational}. Here we modulate electronic structure by using selenium element substitution, which may change the topological properties of ZrTe$_5$. In order to design a large-gap band topological nontrivial phase, one widely used approach is applying strain, which has been proved to be able to regulate the topological properties, such as SiGe\cite{sige}, TlSbH$_2$\cite{Tl}, GaBi(CH$_3$)$_2$ \cite{Ga} and HgSe\cite{hgse}, so we try to use strain to increase the band gap of ZrTe$_4$Se. \\

 \indent In this paper, the first-principles calculations are used to investigate the electronic  properties of the ZrTe$_4$Se.  We find the band structure of bulk ZrTe$_4$Se can produce a semimetal-to-TI phase transition under certain uniaxial strain.  Analyzing of topological properties of bulk ZrTe$_4$Se under the 3\% and 8\% tensile strain, they are all WTIs with Z$_2$ = (0; 110). The single-layer ZrTe$_4$Se is a QSH insulator with Z$_2$ = 1, the edge states further confirm the nontrivial topological nature of this material. In addition, we discuss effect of strain on electronic properties of single-layer ZrTe$_4$Se and find the QSH states survives at a large range of strain. When the compressive strain is more than 1\%, the band structure of single-layer ZrTe$_4$Se can produce a TI-to-semimetal transition.

\section*{II. COMPUTATIONAL DETAILS}
To study the structural and electronic properties of ZrTe$_4$Se, all calculations are carried out using the Vienna ab initio simulation package (VASP) \cite{26,27}. We use the generalized gradient approximation (GGA)\cite{28} for the exchange and correlation potential in the Perdew-Burke-Ernzerhof (PBE)\cite{29} form. The vacuum region is set to at least 20 {\AA} along \textit{b}-direction of the sheets to avoid interactions between neighboring slabs for single-layer material. The energy cutoff of the plane wave is set to 500 eV with the energy precision of 10$^{-6}$ eV. The atoms are relaxed until the force per atom falls below 0.01 eV/{\AA}.  A $11\times11\times4$ ($11\times1\times4$) $\Gamma$-centered Monkhorst-Pack grid\cite{30} for the bulk materials (2D materials) is used to sample the Brillouin zone. The theoretical ground states of ZrTe$_4$Se is obtained by fully optimization of the atom positions and lattice constants, then we vary the lattice constants and optimize the atom positions to study the possible topological transition in ZrTe$_4$Se, but the crystal structure symmetry remains the same. To obtain good theoretical lattice constants, the van der Waals (vdw) corrected optB86b-vdw functionals\cite{31,32} are considered for the bulk materials. What's more, we have checked the electronic structure by using the hybrid functional HSE06\cite{87}, it gives a band gap of 0.46 eV for single-layer ZrTe$_4$Se in Supplemental Material (SM) Fig. S1. To analyze topological properties, the maximally localized Wannier functions (MLWFs)\cite{MLWF1,MLWF2,MLWF3} of ZrTe$_4$Se are constructed based on the Zr's 4d, Se's 4p and Te's 5p orbitals by using the Wannier90 code\cite{MLWF1,MLWF2}. After successful constructions of the MLWFs, the WannierTools\cite{35} is used to evaluate Z$_2$ topological invariants, surface states and edge states.\\
\section*{III. RESULTS AND DISCUSSION}
\begin{figure}[htp!]
\centerline{\includegraphics[width=0.8\textwidth]{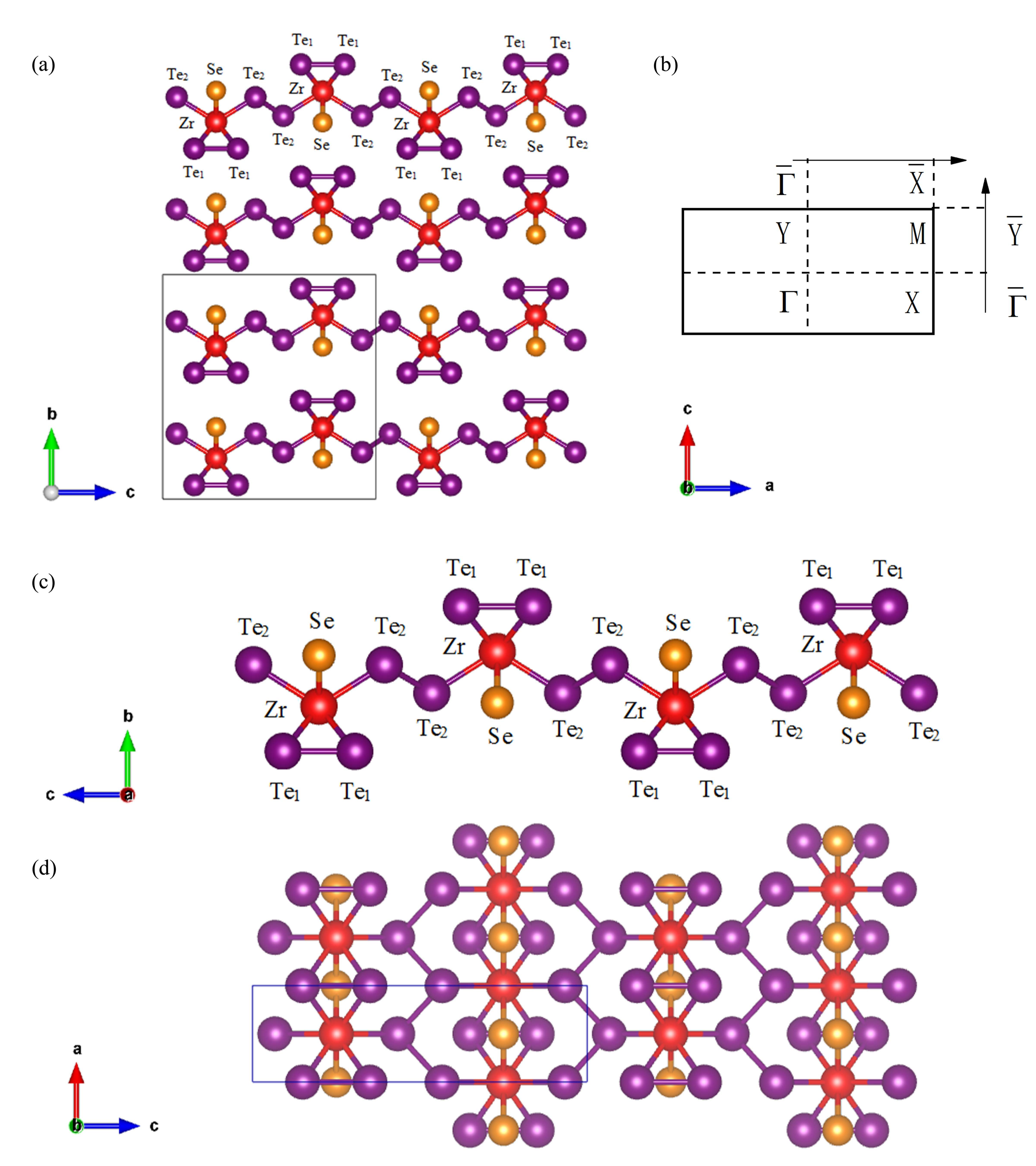}}
\caption{(Color online) The crystal structure and Brillouin zone (BZ) of ZrTe$_4$Se. (a) The crystal structure of bulk ZrTe$_4$Se.  (b) The BZ of single-layer ZrTe$_4$Se. (c) The side and (d) top view the crystal structure of single-layer ZrTe$_4$Se, the vacuum stack along the b direction. }
\label{fig:latt}
\end{figure}

\begin{table*}[!htbp]
\centering
\caption{The lattice constants \textit{a}, \textit{b}, \textit{c} ({\AA}) and bond lengths ({\AA}) of ZrTe$_4$Se.}
\begin{tabular}{c|c|c|c|c|c|c} 
\hline
\hline
 Material&$\textit{a}$&$\textit{b}$&$\textit{c}$&$\textit{d$_{Zr-Te_1}$}$&$\textit{d$_{Zr-Te_2}$}$&$\textit{d$_{Zr-Se}$}$\\
\hline
2D-ZrTe$_4$Se&3.9601&-&13.7411&2.98&2.99&2.79\\
3D-ZrTe$_4$Se&3.9179&14.3565&13.6428&2.96&2.96&2.79\\
\hline
\end{tabular}
\end{table*}

\begin{figure}[htp!]
\centerline{\includegraphics[width=0.8\textwidth]{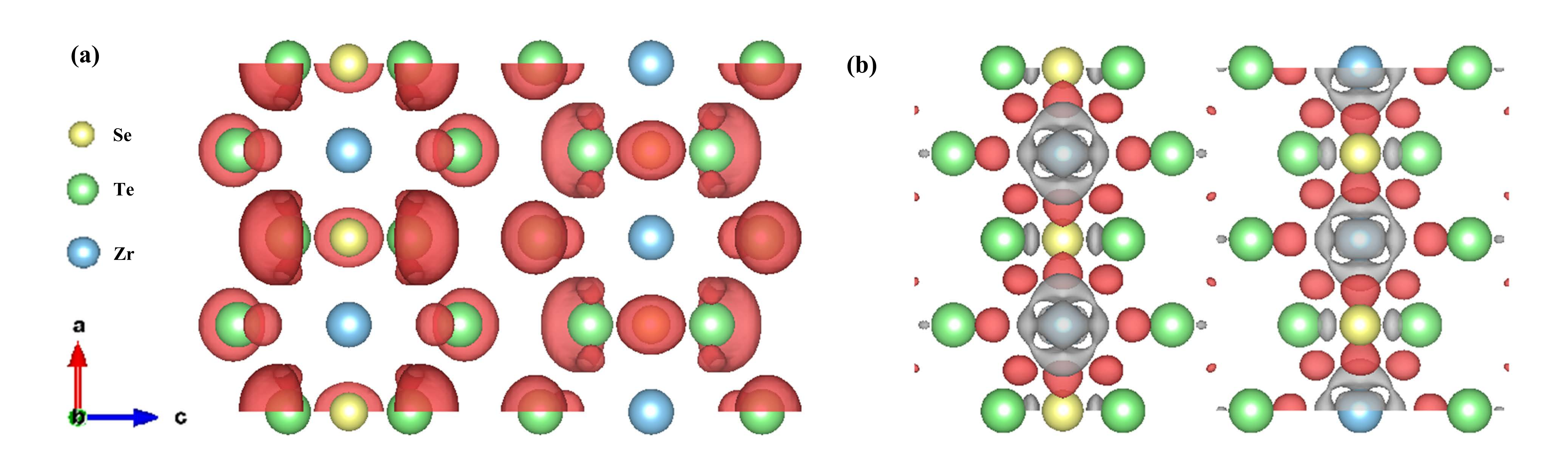}}
\caption{(Color online)  Electron localization function (ELF) and difference charge density of single-layer ZrTe$_5$Se. (a) Structure plot of ELF. Isosurface corresponding to ELF value of 0.88. (b) Difference charge density (crystal density minus superposition of isolated atomic densities). The red (gray) isosurface plots correspond to the charge density accumulation (depletion). Isosurface corresponding to difference charge density of $\pm$ 0.006 eV/${\AA}^3$.}
\label{fig:charge}
\end{figure}

The ZrTe$_4$Se has the orthorhombic layered structure with Cmcm (No. 63) space group symmetry, as shown in Fig. \ref{fig:latt} (a). The trigonal prismatic chains Te$_1$-Se-Te$_1$ oriented along the \textit{a} axis, and these prismatic chains are linked via parallel zigzag chains of Te$_2$ atoms along the $\textit{c}$ axis to form a 2D sheet of ZrTe$_4$Se in the $\textit{a}$-$\textit{c}$ plane. The sheets stack along the \emph{b} axis, forming a layered structure. The structural parameters of  ZrTe$_4$Se are presented in Table I. The optimized lattice constants of bulk ZrTe$_4$Se are $\textit{a}$= 3.9179 {\AA}, $\textit{b}$= 14.3565 {\AA}, $\textit{c}$= 13.6428 {\AA}, which are smaller than ZrTe$_5$'s\cite{36}. The binding energy of -1.70 eV/atom indicates the material is stable. The formation energy of 0.09 eV/atom indicates the single-layer of ZrTe$_4$Se may be produced by mechanical exfoliation method\cite{MEM}. Fig. \ref{fig:latt} (b) - (d) show the Brillouin zone and crystal structure of the single-layer ZrTe$_4$Se. The relaxed lattice constants are 3.9601 {\AA} and 13.7411 {\AA}, the bond lengths are also slightly different from bulk materials, which can be attributed to the interlayer interactions. We use the electron localization function (ELF) to describe and visualize chemical bonds in solids\cite{elf}. The result for single-layer  ZrTe$_4$Se is illustrated in Fig. \ref{fig:charge} (a) for the ELF = 0.88 isosurface. The greater value of Te-Te (or Te-Se) bonding suggest a strong covalent bonding character, while the Zr-Te (or Zr-Se) bonding indicate the highly ionic nature.  To qualitatively analyze the charge transfer of Zr-Te (or Zr-Se) bond, difference charge density map is plotted in Fig. \ref{fig:charge} (b), the red/gray region represents charge accumulation/depletion, respectively. The difference pattern indicates the major charge transfer is from Zr atom to Te (or Se) atom.
In addition, we perform ab initio molecular dynamics (MD) simulation with a supercell at 300 K to examine thermal dynamic stability of ZrTe$_4$Se. After heating at 300 K for 8ps with a time step of 2fs, it is found that the mean value of total potential energy maintains invariable at whole simulation time, see SM Fig. S2. Neither structure reconstruction nor disruption occur in these materials in Fig. S3 and S4. These results clearly indicate the materials remain thermally stable at room temperature.
Moreover, we study mechanical stability by calculating elastic constants of bulk materials. For the orthorhombic crystals, there are nine independent elastic stiffness constants in Table II.  They fulfill the Born criteria of stability\cite{elastic}, C$_{11}$ $>$ 0, C$_{11}$C$_{22}$ $>$ C$_{12}^2$, C$_{11}$C$_{22}$C$_{33}$ + 2C$_{12}$C$_{13}$C$_{23}$ - C$_{11}$C$_{23}^2$ - C$_{22}$C$_{13}^2$ - C$_{33}$C$_{12}^2$ $>$ 0,  C$_{44}$ $>$ 0,  C$_{55}$ $>$ 0 and C$_{66}$ $>$ 0, indicating bulk ZrTe$_5$ and ZrTe$_4$Se are all mechanically stable.

\begin{table*}[!htbp]
\centering
\caption{Elastic constants (GPa) of bulk ZrTe$_4$Se and ZrTe$_5$.}
\begin{tabular}{c|c|c|c|c|c|c|c|c|c} 
\hline
\hline
 Material&C$_{11}$&C$_{12}$&C$_{13}$&C$_{22}$&C$_{23}$&C$_{33}$&C$_{44}$&C$_{55}$&C$_{66}$\\
\hline
3D-ZrTe$_5$&77.20&4.77&22.19&28.85&4.20&70.74&4.92&0.66&30.93\\
3D-ZrTe$_4$Se&90.12&2.92&21.59&28.20&9.30&67.07&1.27&24.43&1.43\\
\hline
\hline
\end{tabular}
\end{table*}

\begin{figure}[htp!]
\centerline{\includegraphics[width=0.8\textwidth]{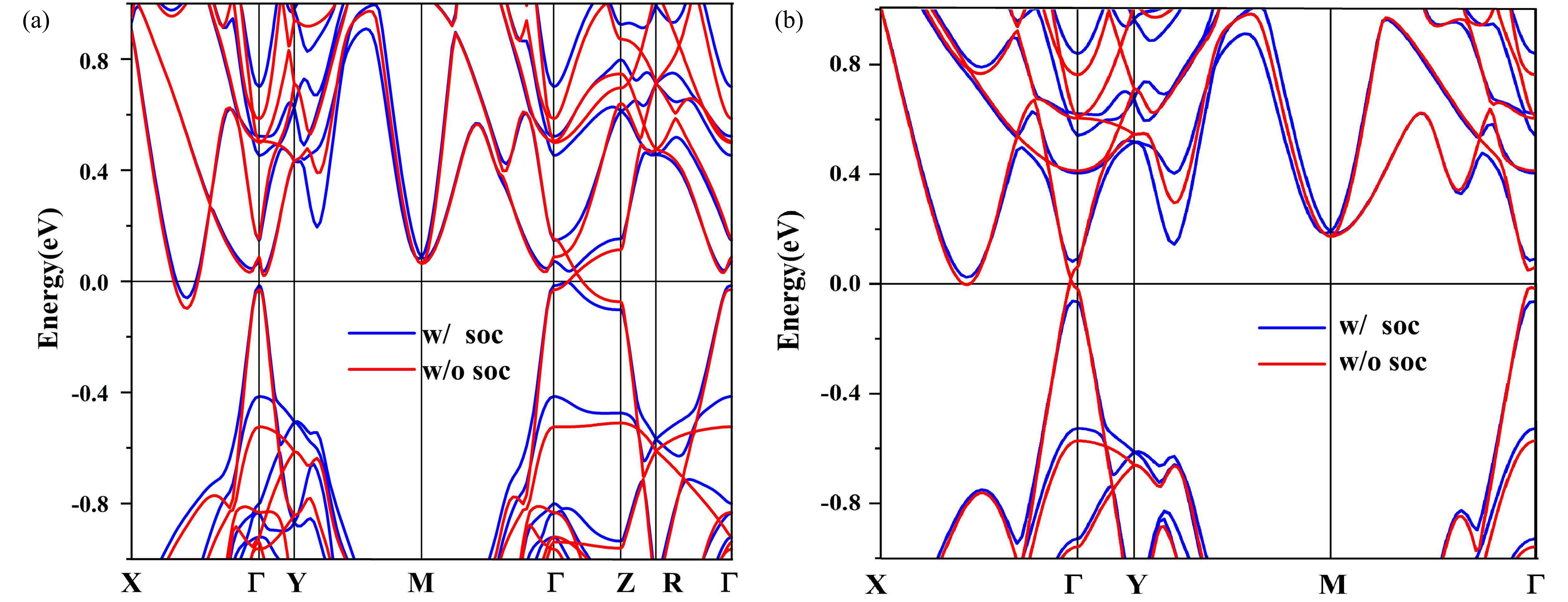}}
\caption{(Color online) The band structures of (a) bulk and (b) single-layer of ZrTe$_4$Se. The red and blue lines correspond to band structures without and with SOC, respectively. The Fermi energy is set to 0 eV. }
\label{fig:band}
\end{figure}

\begin{figure}[htp!]
\centerline{\includegraphics[width=1.1\textwidth]{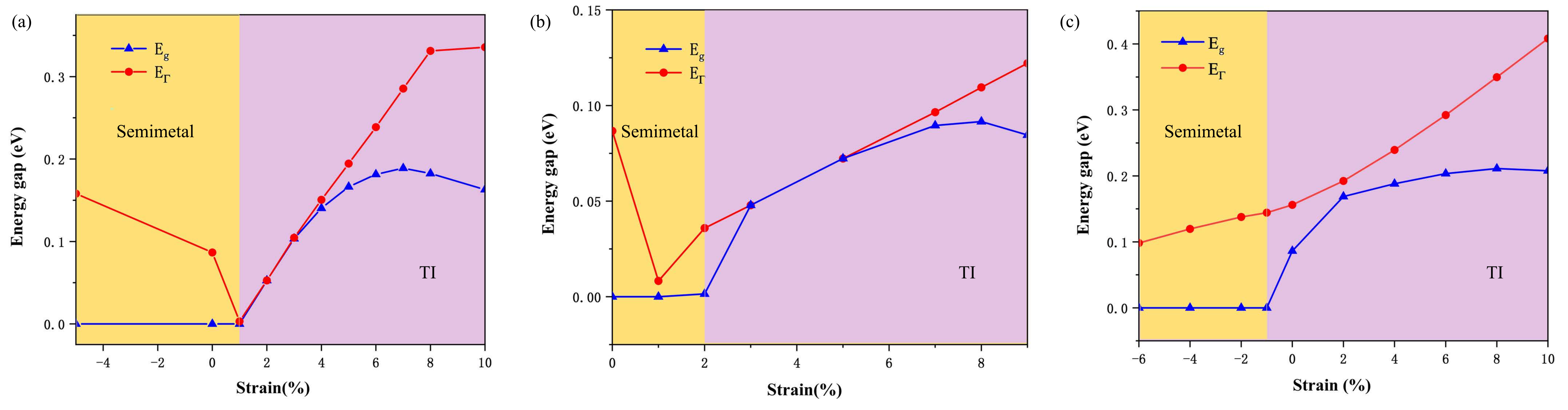}}
\caption{(Color online) The variation of band gap as a function of uniaxial strain along the (a) [100], (b) [111] direction of bulk ZrTe$_4$Se and (c) [100] direction of single-layer ZrTe$_4$Se. E$_g$ and E$_\Gamma$ represent the globe band gap and direct band gap at the $\Gamma$ point, respectively.  The nontrivial Z$_2$ topology survives as long as the globe band gap remains positive. }
\label{fig:strain}
\end{figure}

\begin{figure}[htp!]
\centerline{\includegraphics[width=0.8\textwidth]{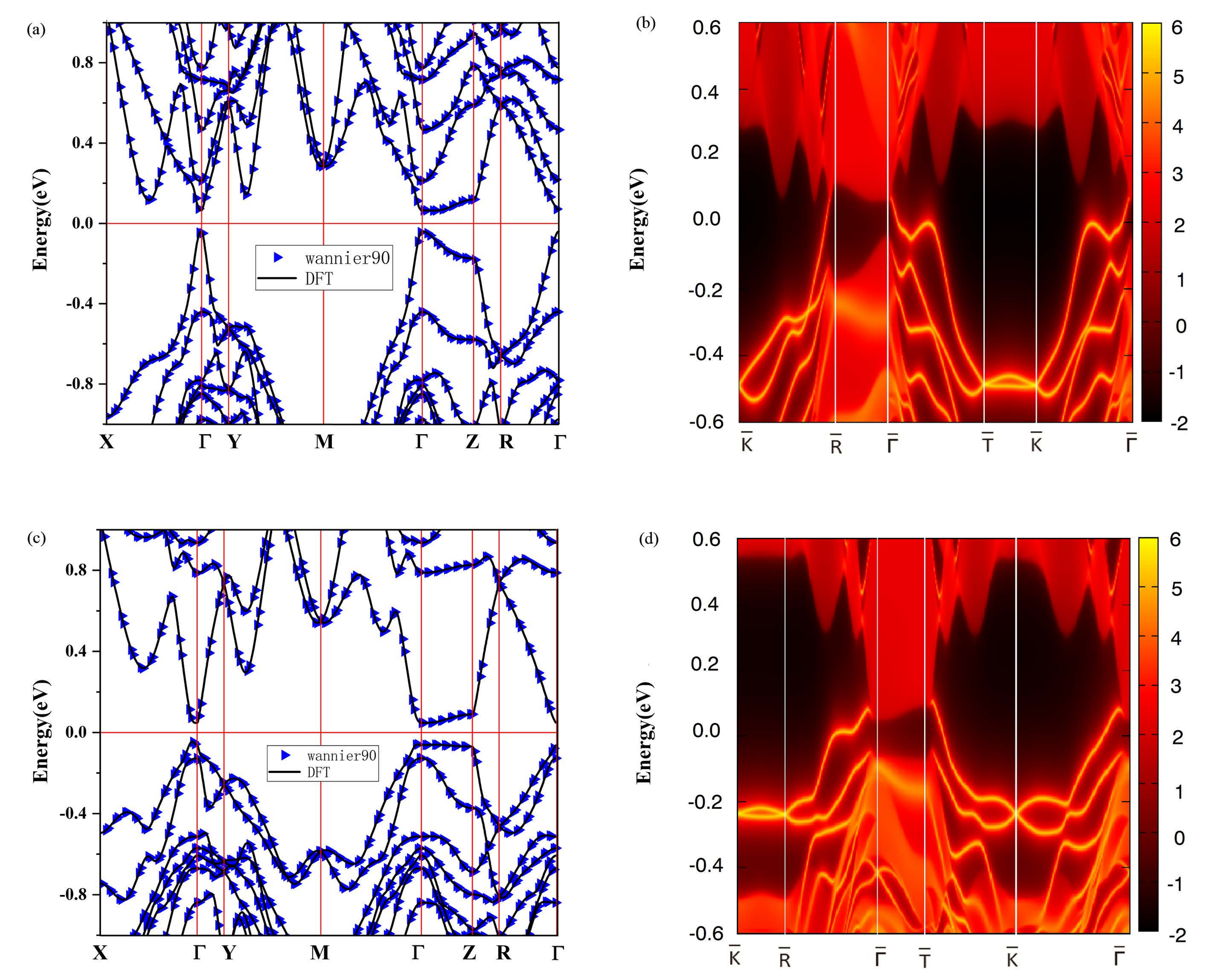}}
\caption{(Color online) The band structures and surface states of bulk ZrTe$_4$Se. The band structures of ZrTe$_4$Se under (a) 3\% [100] and (c)  8\% [111] uniaxial strain. The (001) surface states of ZrTe$_4$Se under  (b) 3\% [100] and (d) 8\% [111] uniaxial strain. }
\label{fig:state}
\end{figure}

The calculated band structures for bulk ZrTe$_4$Se are shown in Fig. \ref{fig:band}  (a). Without SOC, the band structure presents a semimetal.  With the consideration of SOC, the crossing point on the $\Gamma$$-Z$ direction is separated and the conduction band minimum rises, but it has still a semimetal phase.  We study the effect of different strain on electronic and topological properties in the bulk ZrTe$_4$Se and find the uniaxial strain along the [010] and [001] directions is little effect on the electronic properties, so the main research tasks focus on the uniaxial strain along the [100] and [111] directions. The variation of band gap (E$_{\Gamma}$ and E$_g$) as a function of uniaxial strain along the [100] direction is presented  Fig. \ref{fig:strain} (a).  E$_g$ and E$_\Gamma$ represent the globe band gap and direct band gap at the $\Gamma$ point, respectively.  It can be seen that when the tensile strain is more than 1\%, the phase transition from a semimetal to semiconductor. The E$_{\Gamma}$ increases monotonically under strain from 1\% to 10\%, reaching a maximum value of 0.336 eV at 10\% and minimum value of 3.1 meV at 1\%. The E$_g$ increases first then decreases under tensile strain increases continuously, reaching a maximum value of 0.189 eV at 7\%. 
The topological properties of ZrTe$_4$Se have been checked and nontrivial topological phases exists from 1\% to 10\% strain range. We choose ZrTe$_4$Se under [100] strain at 3\% as a typical example, the four Z$_2$ invariants are (0; 110) and it is a WTI  according to method proposed by Fu, Kane and Mele\cite{Fu}. The existence of gapless surface states is an additional commonly employed criterion for TI. The band structure and (001) surface state are shown in Fig. \ref{fig:state}.  The VBM and CBM are separated resulting 0.103 eV band gap. In the exotic topological surface state, two Dirac cones located at the $\overline{T}$ and $\overline{K}$ points respectively for the (001) surface. \\
 \indent Then we study the effect of  uniaxial  strain along the [111] direction, the variation of band gap as a function of strain is presented  Fig. \ref{fig:strain} (b). The E$_{\Gamma}$ increases monotonically under strain from 1\% to 9\%,  the E$_g$ increases with strain and reaches a maximum value of 91.6 meV at 8\%. When the strain is more than 2\%, a semiconductor phase occurs. In addition, we find it is still a WTI with same Z$_2$ under 2\% to 9\% uniaxial strain. The band structure and (001) surface state for ZrTe$_4$Se under 8\% strain are presented in Fig. \ref{fig:state} (c) and (d), two Dirac cones located at the $\overline{R}$ and $\overline{K}$ points confirm it is WTI.\\

\begin{figure}[htp!]
\centerline{\includegraphics[width=0.8\textwidth]{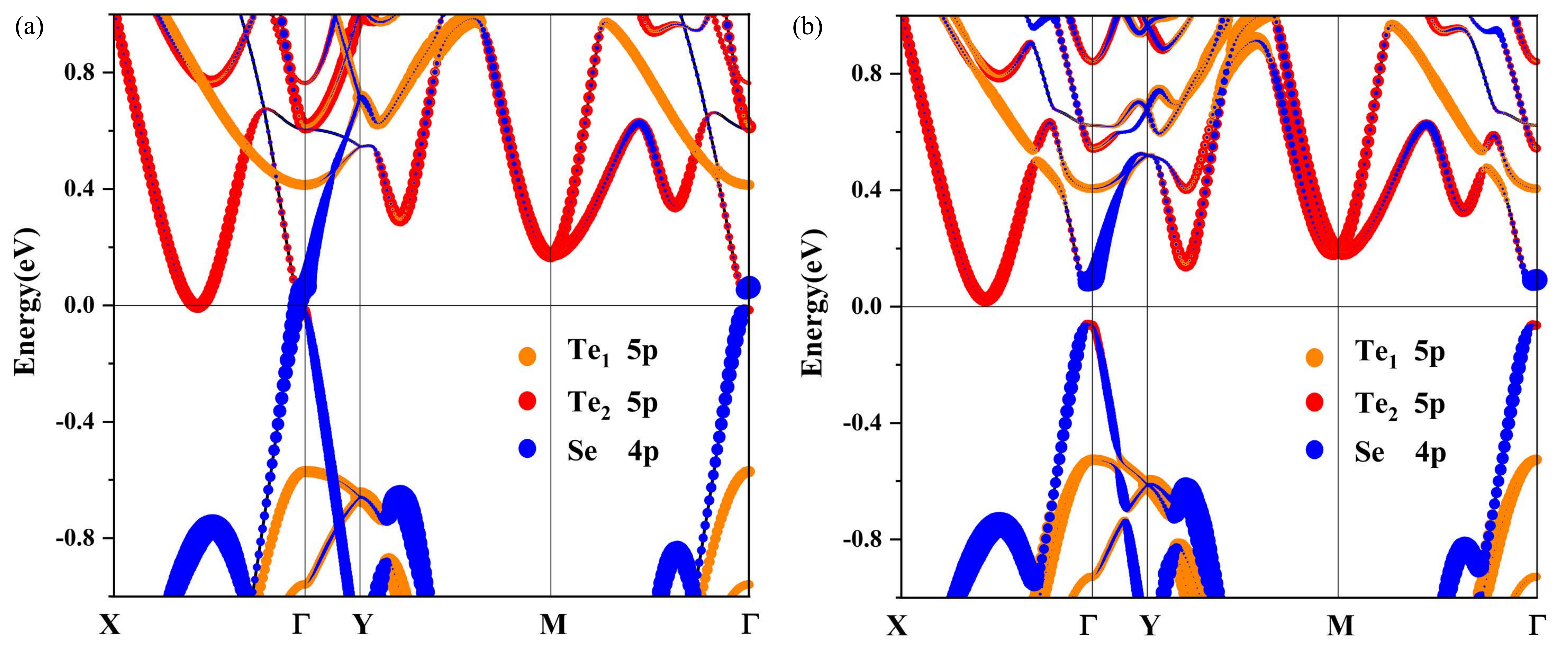}}
\caption{(Color online) The orbitals-resolved band structures of single-layer ZrTe$_4$Se (a) without and (b) with SOC.   }
\label{fig:orbital}
\end{figure}

\begin{figure}[htp!]
\centerline{\includegraphics[width=1.0 \textwidth]{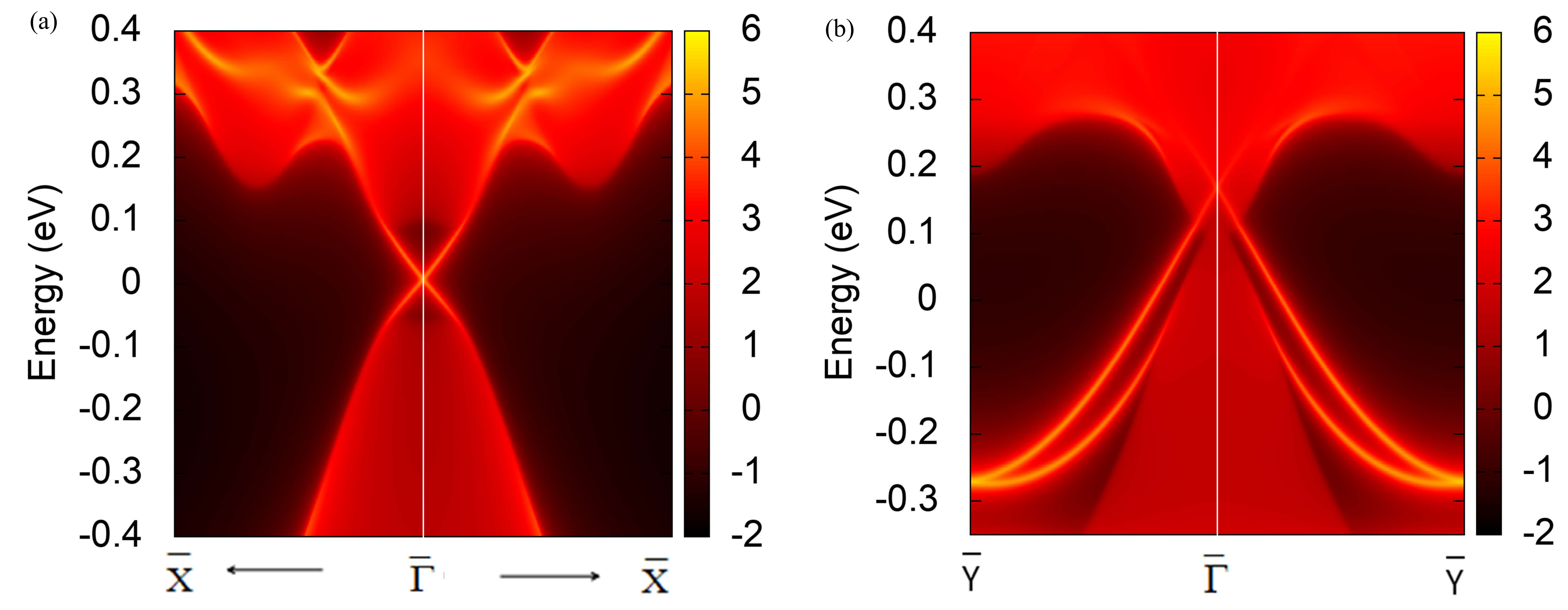}}
\caption{(Color online)  The calculated edge states for (a) \textit{a} axis and (b) \textit{c} axis edges, respectively.}
\label{fig:wcc}
\end{figure}

The calculated band structures for a single-layer ZrTe$_4$Se are displayed in Fig. \ref{fig:band} (b). It is a gapless semimetal with the CBM aross the Fermi energy without consideration of SOC, the VBM and CBM near the $\Gamma$ point almost touch each other. With consideration of SOC, the band structure produces a semimetal-to-semiconductor transition with a band gap of 86.4 meV. By presenting the orbitals-resolved band structures in Fig. \ref{fig:orbital}, it can be seen the bands around the Fermi level are mainly derived from Te$_2$ 5p and Se 4p orbitals.  The evolution lines of Wannier centers in Fig. S5 show it is a nontrivial QSH insulator with Z$_2$=1 and the important character of helical edge states also appear, as shown in Fig. \ref{fig:wcc}. Helical edge states are very useful for electronics and spintronics owing to their robustness against scattering. There is a Dirac cone at the $\overline{\Gamma}$ for the \textit{a} axis edge. Fermi velocity of helical dege states, which is an important quantity related to applications, is about $3.2\times 10^5$ m/s, comparable to $10^6$ m/s in graphene\cite{grap}. For the \textit{c} axis edge, the symmetric edge structure leads to two Dirac cones located at opposite $\overline{Y}$ points. The nontrivial metallic edge states further confirm the nontrivial topological nature of the monolayer ZrTe$_4$Se.\\

\begin{figure}[htp!]
\centerline{\includegraphics[width=0.6 \textwidth]{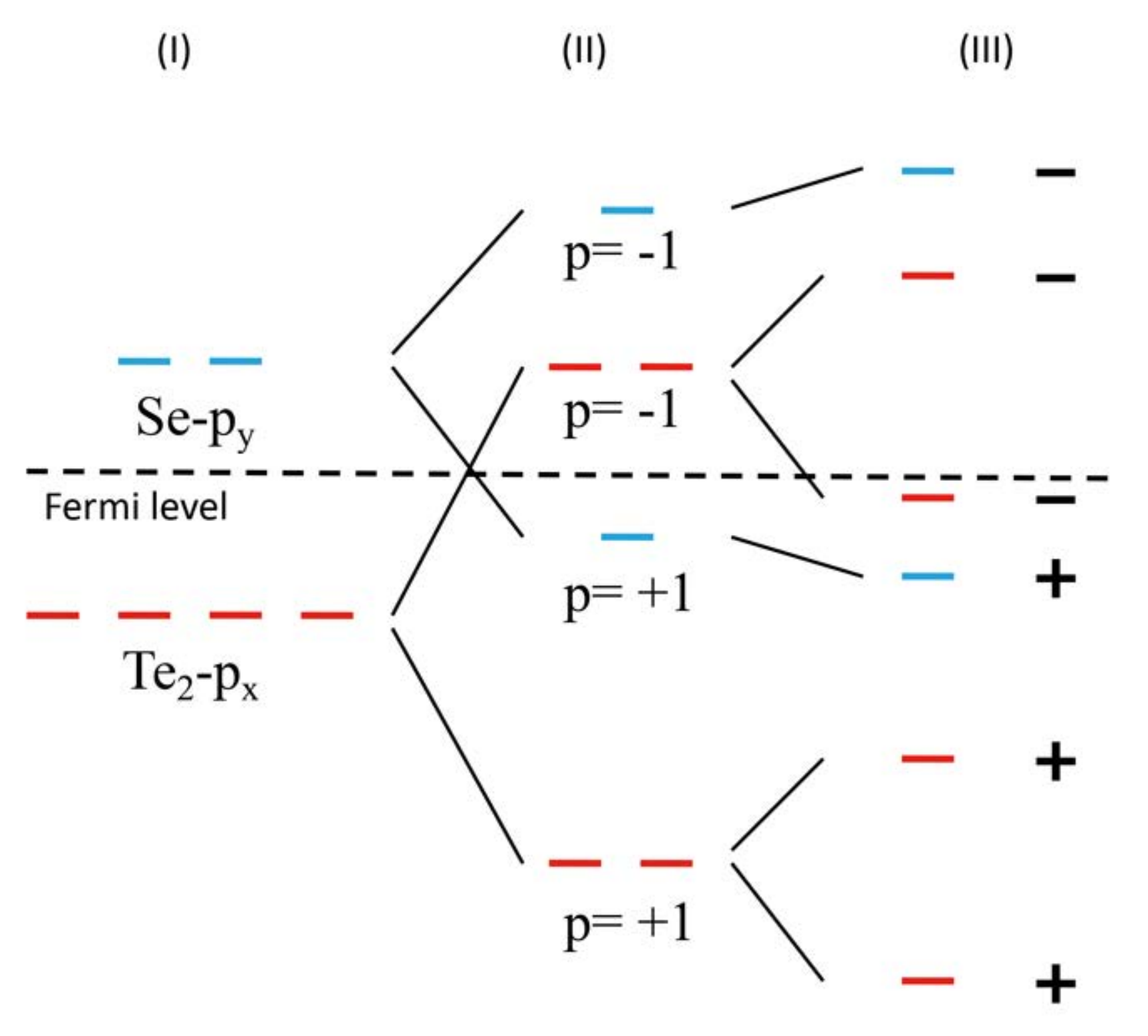}}
\caption{(Color online)  The band inversion mechanism in single-layer ZrTe$_4$Se.}
\label{fig:band-rev}
\end{figure}

To get a physical understanding of the topological nature, we start from atomic orbitals and consider the effect of chemical bonding on the energy levels at the $\Gamma$ point for  monolayer ZrTe$_4$Se. For the convenience of discussion, we define coordinate system with x, y along a, c axes, respectively. The origin of the coordinate system located on Zr site, so the inversion center located at (0.25, 0.25). We note single-layer ZrTe$_4$Se has space group Pmmn ($D_{2h}^{13}$), which is nonsymmorphic, the Z$_2$ index of the material is fully determined by the energy order of the bands at the $\Gamma$ point\cite{zrte1}. From the orbitals-resolved band structures, we find the band inversion happens between the Te$_2$-p$_x$ and Se-p$_y$, as shown in Fig. \ref{fig:band-rev}. For (I) process, there are four equivalent Te$_2$ atoms, they are fourfold degenerate. There are two equivalent Se atoms, they are double degenerate. For (II) process, the strong intrachain covalent bonding will split them into bonding and antibonding states. The Te$_2$ and Se have inversion symmetry and can be divided into two classes with p=+1 or p=-1. For (III) process, the weak interchain coupling will further change the Se's states and split Te$_2$ states to single non-degenerate states. As a result, only the Te$_2$ state has odd parity, the total parity of the occupied states is negative, which leads to the QSH state. The band gap is opened by SOC effect, but SOC effect has nothing to do with topological properties.

\indent  Further more, we study  electronic properties of single-layer ZrTe$_4$Se under different strain to explore the possible phase transition. For the  uniaxial strain along the [100] direction, the variation of band gap as a function of strain is presented Fig. \ref{fig:strain} (c).  The E$_{\Gamma}$ increases monotonically under strain from -6\% to 10\%. The E$_g$ increases under tensile strain increases continuously and reaches a maximum value of 0.211 eV at 8\% tensile strain. It can be seen that the nontrivial topological phases exists over a wide strain from -1\% to 10\%, such robust topology against lattice deformation makes it easier for experimental realization and characterization on different substrate. When the compressive strain is more than 1\%, the band structure produces an TI-to-semimetal transition.
\section*{IV. Conclusion}
In summary, the bulk and single-layer ZrTe$_4$Se is mechanically and dynamically stable, so it is possibly to be prepared. The bulk ZrTe$_4$Se is predicted a new 3D WTI with Z$_2$ invariants (0; 110) under a large range of uniaxial strain. When the tensile strain along the [100] ([111]) direction is more than 1\% (2\%), the band structure of bulk ZrTe$_4$Se produces a semimetal-to-TI transition. The maximum global band gap is 0.189 eV at the  7\% tensile strain along the [100] direction. The two Dirac cones for the (001) surface confirm the nontrivial topological nature at the  [100] ([111]) tensile strain 3\% (8\%). These calculations demonstrate the bulk ZrTe$_4$Se can be turned into a 3D TI via proper strain engineering. The single-layer ZrTe$_4$Se is a QSH insulator with a band gap 86.4 meV and Z$_2$=1, the edge states further confirm the nontrivial topological nature of this material. The Dirac point located at the band gap has a high velocity about $3.2\times 10^5$ m/s. The QSH state survives at a large range of strain from -1\% to 10\%, indicating its robust stability against the strain. The maximum global band gap is 0.211 eV at the tensile strain 8\%. When the compressive strain is more than 1\%, the band structure of single-layer ZrTe$_4$Se produces a TI-to-semimetal transition. These findings make the ZrTe$_4$Se is an excellent candidate for large-gap TI and may provide a platform for realizing low-dissipation quantum spintronic devices.

\begin{acknowledgments}
This work was supported by National Natural Science Foundation of China (No.11904312 and 11904313), the Project of Department of Education of Hebei Province, China(No.BJ2020015), and the Natural Science Foundation of Hebei Province (No. A2019203507 and A2020203027). The authors thank the High Performance Computing Center of Yanshan University.
\end{acknowledgments}


\nocite{*}
\bibliography{zrte4se.bib}

\end{document}